\documentclass[12pt]{article}
 \usepackage[cp1251]{inputenc} 
 \usepackage[english, russian]{babel} 
 \usepackage{amsmath, latexsym, amssymb, bm, array, graphics, eucal,
 amsfonts,}
 \makeatletter
 \renewcommand{\@biblabel}[1]{#1.\hfill}
 \newcommand{\diag}{\rm \diag\, }
 
 \newcommand{\const}{\mathop{\rm const\, }}
 \renewcommand{\Re}{\mathop{\rm Re}}
 \renewcommand{\Im}{\mathop{\rm Im\,}}

\renewcommand{\div}{\mathop{\rm div}}
\newcommand{\mc}[1]{\mathcal{#1}}
\newcommand{\E}{\mc{E}}
 \makeatother
 \renewcommand{\baselinestretch}{1.025}
\usepackage[dvips]{graphicx}
\usepackage[flushleft,small]{caption2} 

\begin{document}
 \thispagestyle{empty}
\large
\renewcommand{\refname}{\center ЛИТЕРАТУРА}
\renewcommand{\abstractname}{Abstract }
 \begin{center}
\bf Transversal electric conductivity and dielectric function in quantum 
Maxwell collisional plasma 
\end{center}
\medskip
\begin{center}
  \bf A. V. Latyshev\footnote{$avlatyshev@mail.ru$} and
  A. A. Yushkanov\footnote{$yushkanov@inbox.ru$}
\end{center}\medskip

\begin{center}
{\it Faculty of Physics and Mathematics,\\ Moscow State Regional
University, 105005,\\ Moscow, Radio str., 10--A}
\end{center}\medskip

\begin{abstract}
Formulas for transversal conductance and dielectric permeability (dielectric
function) in
quantum Maxwell  collisional plasma are deduced. The kinetic
equation with collision integral in the form relaxation type is used. 


\end{abstract}

\begin{center}\bf
   ВВЕДЕНИЕ
\end{center}

В настоящей работе выводятся формулы для вычисления электрической проводимости
и диэлектрической проницаемости в квантовой
максвелловской столкновительной плазме.
При выводе кинетического уравнения нами обобщается подход,
развитый Климонтовичем и Силиным \cite{Klim}.

Диэлектрическая проницаемость в бесстолкновительной квантовой
газовой плазме изучалась многими авторами \cite{Klim}--\cite{Brod}. 
В работе \cite{Manf}, где
исследован одномерный случай квантовой
плазмы, отмечалась важность вывода диэлектрической проницаемости с
использованием квантового кинетического уравнения с интегралом
столкновений в форме БГК--модели (Бхатнагар, Гросс, Крук).
Настоящая работа посвящена выполнению этой задачи для максвелловской плазмы.

Диэлектрическая проницаемость является одной из важнейших
характеристик плазмы. Эта величина необходима для анализа поверхностных 
плазмонов \cite{Fuchs}, для изучения механизма проникновения
электромагнитных волн в плазму \cite{Shukla1}, и для
анализа других проблем в физике плазмы.

Кливер и Фукс \cite{Kliewer} первыми заметили, что выведенная
Линдхардом диэлектрическая функция для квантовой плазмы
в столкновительном режиме не переходит
в диэлектрическую функцию для классической плазмы в пределе, когда
постоянная Планка $\hbar$ стремится к нулю.
Это значит, что диэлектрическая функция Линдхарда не учитывает
корректно столкновения электронов.
Кливер и Фукс "подправили"\, диэлектрическую функцию Линдхарда
так, чтобы она переходила в классическую при $\hbar \to 0$.
В работе \cite{Mermin} был проведен корректный учет столкновений
в рамках релаксационной модели в пространстве импульсов
электронов при построении продольной диэлектрической функции.

В то же время для поперечной диэлектрической проницаемости
корректного учета влияния столкновений в квантовой максвелловской 
плазме до сих пор проведено не было.

Целью настоящей работы является устранение указанного выше
пробела и вывод формул электрической проводимости и диэлектрической функции
для квантовой максвелловской столкновительной плазмы.

\begin{center}
  {\bf 1. КИНЕТИЧЕСКОЕ УРАВНЕНИЕ ДЛЯ ФУНКЦИИ ВИГНЕРА}
\end{center}

Рассмотрим уравнение Шредингера, записанное
для частицы в электромагнитном поле на матрицу плотности $\rho$:

$$
i\hbar \dfrac{\partial \rho}{\partial t}=H\rho-{H^*}'\rho.
\eqno{(1.1)}
$$

Здесь $H$ -- оператор Гамильтона, $H^*$ -- комплексно сопряженный к
$H$ оператор, ${H^*}'$ -- комплексно сопряженный к
$H$ оператор, действующий на штрихованные пространственные переменные
$\mathbf{r}'$.

Оператор Гамильтона свободной частицы, находящейся в поле скалярного
потенциала $U$ и в поле векторного потенциала $\mathbf{A}$,
имеет вид:
$$
H=\dfrac{(\mathbf{p}-\dfrac{e}{c}\mathbf{A})^2}{2m}+eU=$$$$=
\dfrac{\mathbf{p}^2}{2m}-
\dfrac{e}{2mc}(\mathbf{p}\mathbf{A}+\mathbf{A}\mathbf{p})
+\dfrac{e^2}{2mc^2}\mathbf{A}^2+eU.
\eqno{(1.2)}
$$

Здесь $\mathbf{p}$ -- оператор импульса, $\mathbf{p}=-i\hbar \nabla$,
$e$ -- заряд электрона, $m$ -- его масса, $c$ -- скорость света.

Операторы $H$ и ${H^*}'$, вычисленные от матрицы плотности, имеют вид:
$$
H\rho=-\dfrac{\hbar^2}{2m}\Delta \rho+\dfrac{ie\hbar}{2mc}
\Big(2\mathbf{A}\nabla \rho+\rho\nabla\mathbf{A}\Big)+
\dfrac{e^2}{2mc^2}\mathbf{A}^2\rho+eU \rho
\eqno{(1.3)}
$$
и
$$
{H^*}'\rho=-\dfrac{\hbar^2}{2m}\Delta'\rho-\dfrac{ie\hbar}{2mc}
\Big(2\mathbf{A'}\nabla'\rho+\rho\nabla'\mathbf{A}\Big)+
\dfrac{e^2}{2mc^2}\mathbf{A'}^2\rho+eU' \rho.
\eqno{(1.4)}
$$

Операторы $\nabla$ и $\Delta$ в (1.3) и (1.4) действуют на
нештрихованные пространственные переменные матрицы плотности, т.е.
$\nabla=\nabla_{\mathbf{R}}$, $\Delta=\Delta_{\mathbf{R}}$.
В операторе ${H^*}'$ следует заменить операторы
$\nabla=\nabla_{\mathbf{R}}$ и $\Delta=\Delta_{\mathbf{R}}$
на операторы
$\nabla'\equiv\nabla_{\mathbf{R}'}$ и
$\Delta'\equiv\Delta_{\mathbf{R}'}$,
кроме того, введены обозначения
$
\mathbf{A'}\equiv \mathbf{A}(\mathbf{R'},t),
\quad U'\equiv U(\mathbf{R'},t).
$ 
Связь между матрицей плотности $\rho(\mathbf{r},\mathbf{r}',t)$ и
функцией Вигнера $f(\mathbf{r},\mathbf{p},t)$
дается обратным и прямым преобразованиями Фурье
$$
f(\mathbf{r},\mathbf{p},t)=\int
\rho(\mathbf{r}+\dfrac{\mathbf{a}}{2},\mathbf{r}-
\dfrac{\mathbf{a}}{2},t)e^{-i\mathbf{p}\mathbf{a}/\hbar}d^3a,
$$
$$
\rho(\mathbf{R},\mathbf{R}',t)=\dfrac{1}{(2\pi \hbar)^3}
\int f(\dfrac{\mathbf{R}+\mathbf{R}'}{2}, \mathbf{p},t)
e^{i\mathbf{p}(\mathbf{R}-\mathbf{R}')/\hbar}d^3p.
$$

Функция Вигнера является аналогом функции распределения для квантовых
систем. 
Подставляя представление матрицы плотности через функцию Вигнера
в уравнение Шредингера на матрицу плотности (1.1), получаем
следующее уравнение:
$$
i\hbar \dfrac{\partial \rho}{\partial t}=
\dfrac{1}{(2\pi \hbar)^3}\int\Big\{-\dfrac{i\hbar}{m}
\mathbf{p'}\nabla f +
\dfrac{ie\hbar}{2mc} \Big[\div{\mathbf{A}(\mathbf{R},t)}+
\div{\mathbf{A}(\mathbf{R}',t)}\Big]f+
$$
$$
+\dfrac{ie\hbar}{2mc}\Big[\mathbf{A}(\mathbf{R},t)+
\mathbf{A}(\mathbf{R}',t)\Big]
\nabla f-\dfrac{e}{mc}\Big[\mathbf{A}(\mathbf{R},t)-
\mathbf{A}(\mathbf{R}',t)\Big]
\mathbf{p'}f+
$$
$$
+\dfrac{e^2}{2mc^2}\big[\mathbf{A}^2(\mathbf{R},t)-
\mathbf{A}^2(\mathbf{R}',t)\big]f+
e\big[U(\mathbf{R},t)-U(\mathbf{R}',t)\big]f\Big\}
e^{i\mathbf{p'}(\mathbf{R}-\mathbf{R}')/\hbar}
d^3p'.
\eqno{(1.5)}
$$

В уравнении (1.5) положим
$$
\mathbf{R}=\mathbf{r}+\dfrac{\mathbf{a}}{2},\qquad
\mathbf{R}'=\mathbf{r}-\dfrac{\mathbf{a}}{2}.
$$

Тогда в этом уравнении
$$
 f\Big(\dfrac{\mathbf{R}+\mathbf{R}'}{2}, \mathbf{p'},t\Big)
e^{i\mathbf{p'}(\mathbf{R}-\mathbf{R}')/\hbar}=f(\mathbf{r}, \mathbf{p'},t)
e^{i\mathbf{p'\,a}/\hbar}.
$$

Опуская длительные вычисления, приведем кинетическое уравнение для функции
Вигнера:
$$
\dfrac{\partial f}{\partial t}+\dfrac{1}{m}
\Big(\mathbf{p}-\dfrac{e}{c}\mathbf{A}\Big)\nabla f-\dfrac{e}{mc}
(\div{\mathbf{A}(\mathbf{r},t)})f(\mathbf{r},\mathbf{p},t)=
W[f].
\eqno{(1.6)}
$$
                                                          
В уравнении (1.6) символ $W[f]$ -- интеграл Вигнера --- Власова, 
определяемый равенством
$$
W[f]=
\iint\left\{
\dfrac{e}{2mc}
\Big[\mathbf{A}(\mathbf{r}+\dfrac{\mathbf{a}}{2},t)+
\mathbf{A}(\mathbf{r}-\dfrac{\mathbf{a}}{2},t)-
2\mathbf{A}(\mathbf{r},t)\Big]\nabla f\right.+
$$\smallskip
$$
+\dfrac{ie}{ mc\hbar}\Big[
\mathbf{A}(\mathbf{r}+\dfrac{\mathbf{a}}{2},t)
-\mathbf{A}(\mathbf{r}-\dfrac{\mathbf{a}}{2},t)\Big]\mathbf{p'}
f+
$$\smallskip
$$
+\dfrac{e}{2mc}\Big[\div{\mathbf{A}(\mathbf{r}+
\dfrac{\mathbf{a}}{2},t)
}+\div{\mathbf{A}(\mathbf{r}-\dfrac{\mathbf{a}}{2},t)}-
2\div{\mathbf{A}(\mathbf{r},t)}\Big]f-
$$\smallskip
$$-
\dfrac{i e^2}{2 mc^2\hbar}\Big[\mathbf{A}^2(\mathbf{r}+
\dfrac{\mathbf{a}}{2},t)-
\mathbf{A}^2(\mathbf{r}-\dfrac{\mathbf{a}}{2},t)\Big]f-
$$\smallskip
$$
-\left. \dfrac{ie}{\hbar}\Big[U(\mathbf{r}+\dfrac{\mathbf{a}}{2},t)-
U(\mathbf{r}-\dfrac{\mathbf{a}}{2},t)\Big]f\right\}
e^{i(\mathbf{p'}-\mathbf{p})\mathbf{a}/\hbar}
\dfrac{d^3a\,d^3p'}{(2\pi\hbar)^3}.
\eqno{(1.7)}
$$

Энергия частицы равна
$$
\E=\E(\mathbf{r}, \mathbf{p},t)=\dfrac{1}{2m}\Big(\mathbf{p}-
\dfrac{e}{c}\mathbf{A}(\mathbf{r},t)\Big)^2+eU(\mathbf{r},t).
$$

Тогда скорость частицы $\mathbf{v}$ равна
$$
\mathbf{v}=\mathbf{v}(\mathbf{r}, \mathbf{p},t)=
\dfrac{\partial \E}{\partial \mathbf{p}}=
\dfrac{1}{m}\Big(\mathbf{p}-\dfrac{e}{c}\mathbf{A}\Big),
$$
кроме того,
$$
\nabla \mathbf{v}=-\dfrac{e}{mc}\div{\mathbf{A}}.
$$
Следовательно, левая часть уравнения (1.7) равна:
$$
\dfrac{\partial f}{\partial t}+\dfrac{1}{m}
\Big(\mathbf{p}-\dfrac{e}{c}\mathbf{A}\Big)\nabla f-f\dfrac{e}{mc}
\div{\mathbf{A}}=
\dfrac{\partial f}{\partial t}+\mathbf{v}\nabla f+f\nabla \mathbf{A}=
$$
$$
=\dfrac{\partial f}{\partial t}+\nabla(\mathbf{v}f).
$$

Поэтому уравнение (1.7) можно переписать в виде
стандартном для теории переноса виде:
$$
\dfrac{\partial f}{\partial t}+\dfrac{\partial}{\partial \mathbf{r}} 
(\mathbf{v}f)=W[f].
\eqno{(1.8)}
$$

\begin{center}
{\bf 2. РЕЛАКСАЦИОННАЯ МОДЕЛЬ КИНЕТИЧЕСКОГО УРАВНЕНИЯ}
\end{center}

В случае столкновительной плазмы в правую часть кинетического уравнения
(1.8) добавим интеграл столкновений релаксационного типа $$
\dfrac{\partial f}{\partial t}+\dfrac{\partial}{\partial \mathbf{r}}
(\mathbf{v}f)=\dfrac{f^{(0)}-f}{\tau}+
W[f].
\eqno{(2.1)}
$$

В уравнении (2.1) $\tau$ -- среднее время между двумя последовательными
столкновениями, $\tau=1/\nu$, $\nu$ -- эффективная частота столкновений,
$f^{(0)}$ --- локально равновесное распределение Максвелла---Больцмана,
$$
f^{(0)}=N\Big(\dfrac{\beta}{\pi}\Big)^{3/2}e^{-\beta v^2}.
$$

Здесь $k_B$ -- постоянная Больцмана, $T$ -- температура плазмы,
$N$ -- концентрация (числовая плотность) электронов в равновесном состоянии. 
Введем безразмерную скорость электронов $\mathbf{C}(\mathbf{r},\mathbf{p},t)$:
$$
\mathbf{C}(\mathbf{r},\mathbf{p},t)=\dfrac{\mathbf{v}(\mathbf{r},
\mathbf{p},t)}{v_T}=
\dfrac{\mathbf{p}}{p_T}-\dfrac{e}{cp_T}\mathbf{A}(\mathbf{r},t),
$$
где
$v_T=\dfrac{1}{\sqrt{\beta}}$ -- тепловая скорость электронов,
$\beta=\dfrac{m}{2k_BT}$.

Теперь локально равновесная функция может быть представлена
через скорость электронов как
$$
f^{(0)}(\mathbf{r},\mathbf{p},t)=N\Big(\dfrac{\beta}{\pi}\Big)^{3/2}e^{-C^2},
\eqno{(2.2)}
$$
или, полагая $\mathbf{P}=\dfrac{\mathbf{p}}{p_T}$, в явном виде
$$
f^{(0)}(\mathbf{r},\mathbf{P},t)=N\Big(\dfrac{\beta}{\pi}\Big)^{3/2}
\exp\Big[-\Big(\mathbf{P}-\dfrac{e}{cp_T}\mathbf{A}(\mathbf{r},t)\Big)^2\Big].
\eqno{(2.2')}
$$

Отметим, что в случае постоянного потенциала $\mathbf{A}=\const$ равновесная
функция распределения (2.2) является решением уравнения (2.1).

Найдем среднюю
скорость электронов в равновесном состоянии:
$$
\mathbf{u}^{(0)}(\mathbf{r},t)=\dfrac{1}{N(\mathbf{r},t)}
\int \mathbf{v}(\mathbf{r},\mathbf{p},t)
f^{(0)}(\mathbf{r},\mathbf{p},t)d^3v.
$$

После замены переменных
$\mathbf{p}-(e/c)\mathbf{A}= \mathbf{p}'$ получаем:
$
\mathbf{u}^{(0)}=0.
$

Итак, скорость электронов в равновесном состоянии равна нулю.

Заметим, что числовая плотность электронов и их средняя скорость
удовлетворяют обычному уравнению непрерывности:
$$
\dfrac{\partial N}{\partial t}+\mathbf{div}(N\mathbf{u})=0.
\eqno{(2.3)}
$$

Для вывода уравнения непрерывности (2.3) нужно
проинтегрировать кинетическое
уравнение (2.1) по скоростям и
использовать определение числовой плотности и средней скорости.
Затем следует воспользоваться законом сохранения числа частиц и
проверить, что интеграл по скоростям от интеграла
Вигнера---Власова равен нулю. В самом деле, имеем:
$$
\int W[f]\dfrac{2\;d^3p}{(2\pi\hbar)^3}=2\int\int
\Big\{\cdots\Big\}
e^{i\mathbf{p'}\mathbf{a}/\hbar}\delta(\mathbf{a})\,d^3a\,d^3p'=
$$
$$
=2\int \Big\{\cdots\Big\}\Bigg|_{\mathbf{a}=0}d^3p'\equiv 0,
$$
ибо, как нетрудно проверить простой подстановкой,
$$
 \Big\{\cdots\Big\}\Bigg|_{\mathbf{a}=0}\equiv 0.
$$

Здесь символ $\{\cdots\}$ означает то же самое выражение, что и в
правой части соотношения (1.7).

Заметим, что левая часть кинетического уравнения (1.11) или (2.1)
приобретает стандартный для теории переноса вид при следующем
условии калибровки:
$$
\div{\mathbf{A}(\mathbf{r},t)}=0.
\eqno{(2.4)}
$$

При этом, т.е. в случае калибровки (2.4),
кинетическое уравнение (1.11) принимает вид:
$$
\dfrac{\partial f}{\partial t}+\mathbf{v}\dfrac{\partial f}{\partial\mathbf{r}}=B[f,f]+
W[f],
\eqno{(2.5)}
$$
в котором интеграл Вигнера---Власова равен:
$$
W[f]=
\iint\left\{
\dfrac{e}{2mc}
\Big[\mathbf{A}(\mathbf{r}+\dfrac{\mathbf{a}}{2},t)+
\mathbf{A}(\mathbf{r}-\dfrac{\mathbf{a}}{2},t)-
2\mathbf{A}(\mathbf{r},t)\Big]\nabla f\right.+
$$
\hspace{0.5cm}
$$
+\dfrac{ie}{ mc\hbar}\Big[
\mathbf{A}(\mathbf{r}+\dfrac{\mathbf{a}}{2},t)
-\mathbf{A}(\mathbf{r}-\dfrac{\mathbf{a}}{2},t)\Big]\mathbf{p'}
f-
$$\smallskip
$$
-\dfrac{i e^2}{2 mc^2\hbar}
\Big[\mathbf{A}^2(\mathbf{r}+\dfrac{\mathbf{a}}{2},t)-
\mathbf{A}^2(\mathbf{r}-\dfrac{\mathbf{a}}{2},t)\Big]f-
$$
\hspace{0.5cm}
$$
-\left. \dfrac{ie}{\hbar}\Big[U(\mathbf{r}+\dfrac{\mathbf{a}}{2},t)-
U(\mathbf{r}-\dfrac{\mathbf{a}}{2},t)\Big]f\right\}
e^{i(\mathbf{p'}-\mathbf{p})\mathbf{a}/\hbar}
\dfrac{d^3a\,d^3p'}{(2\pi\hbar)^3}.
\eqno{(2.6)}
$$\medskip

\begin{center}
  {\bf 3. ЛИНЕАРИЗАЦИЯ КИНЕТИЧЕСКОГО УРАВНЕНИЯ И ЕГО РЕШЕНИЕ}
\end{center}

Будем рассматривать кинетическое уравнение с интегралом столкновений в
форме $\tau$--модели.

Векторный потенциал возьмем ортогональным
направлению волнового вектора $\mathbf{k}: \mathbf{k}\mathbf{A}=0$
в виде бегущей гармонической волны:
$$
\mathbf{A}(\mathbf{r},t)=\mathbf{A}_0
e^{i(\mathbf{k} \mathbf{r}-\omega t)}.
$$

Будем считать поле векторного потенциала достаточно малым. Это
предположение позволяет линеаризовать уравнение
и пренебречь квадратичными по полю слагаемыми.

Тогда уравнение (2.5) упрощается:
$$
\dfrac{\partial f}{\partial t}+\mathbf{v}\dfrac{\partial f}{\partial \mathbf{r}}=
\dfrac{f^{(0)}-f}{\tau}+W[f].
\eqno{(3.1)}
$$

В уравнении (3.1) локально равновесное распределение Максвелла---Больцмана имеет вид (2.2).
Интеграл Вигнера --- Власова (2.6) также существенно упрощается
и имеет следующий вид:
$$
W[f]=\dfrac{ie}{mc\hbar }
\iint
\Big[\mathbf{A}(\mathbf{r}+\dfrac{\mathbf{a}}{2},t)
-\mathbf{A}(\mathbf{r}-\dfrac{\mathbf{a}}{2},t)\Big]\mathbf{p'}\times
$$
$$
\times
f^{}
e^{i(\mathbf{p'}-\mathbf{p})\mathbf{a}/\hbar}
{d^3a\,d^3v'}.
\eqno{(3.2)}
$$

Заметим, что
$$
\mathbf{A}(\mathbf{r}+\dfrac{\mathbf{a}}{2},t)
-\mathbf{A}(\mathbf{r}-\dfrac{\mathbf{a}}{2},t)=\mathbf{A}(\mathbf{r},t)
\Big[e^{i\mathbf{k}\mathbf{a}/2}-e^{-i\mathbf{k}\mathbf{a}/2}\Big].
$$

Вычисляя интеграл в (3.2), находим, что
$$
W[f]=\dfrac{ie}{mc\hbar}\mathbf{A}(\mathbf{r},t)
\iint \Big[e^{i\mathbf{k}\mathbf{a}/2}-e^{-i\mathbf{k}\mathbf{a}/2}\Big]
e^{i(\mathbf{p}'-\mathbf{p})\mathbf{a}/\hbar}{d^3a\,d^3v'}.
$$

Внутренний интеграл равен:
$$
\dfrac{1}{(2\pi\hbar)^3}\int \Big\{\exp\Big(i\Big[\mathbf{p}'-\mathbf{p}+
\frac{\mathbf{k}\hbar}{2}\Big]\dfrac{\mathbf{a}}{\hbar}\Big)-
\exp\Big(i\Big[\mathbf{p}'-\mathbf{p}-
\frac{\mathbf{k}\hbar}{2}\Big]\dfrac{\mathbf{a}}{\hbar}\Big)\Big\}d^3a=
$$
$$
=\delta(\mathbf{p}'-\mathbf{p}+\dfrac{\hbar\mathbf{k}}{2})-
\delta(\mathbf{p}'-\mathbf{p}-\dfrac{\hbar\mathbf{k}}{2}).
$$

После длительных вычислений находим интеграл Вигнера --- Власова:
$$
W[f]=
$$
$$=\mathbf{A}(\mathbf{r},t)\dfrac{ie}{mc\hbar}
\int \Big[\delta(\mathbf{p}'-\mathbf{p}+\dfrac{\hbar\mathbf{k}}{2})-
\delta(\mathbf{p}'-\mathbf{p}-\dfrac{\hbar\mathbf{k}}{2})\Big]
\mathbf{p}'f(\mathbf{r},\mathbf{p}',t)\,d^3p'=
$$\vspace{0.3cm}
$$
=\mathbf{A}(\mathbf{r},t)\dfrac{ie}{mc\hbar}
\Big[\Big(\mathbf{p}-\dfrac{\hbar \mathbf{k}}{2}\Big)f(\mathbf{r},
\mathbf{p}-\dfrac{\hbar \mathbf{k}}{2},t)-
\Big(\mathbf{p}+\dfrac{\hbar \mathbf{k}}{2}\Big)f(\mathbf{r},
\mathbf{p}+\dfrac{\hbar \mathbf{k}}{2},t)\Big]=
$$\vspace{0.3cm}
$$
=\mathbf{A}(\mathbf{r},t)\dfrac{ie}{mc\hbar}
\Big\{\mathbf{p}\Big[f(\mathbf{r},
\mathbf{p}-\dfrac{\hbar \mathbf{k}}{2},t)-f(\mathbf{r},
\mathbf{p}+\dfrac{\hbar \mathbf{k}}{2},t)\Big]-
$$
\vspace{0.3cm}
$$
-\dfrac{\hbar\mathbf{k}}{2}\Big[f(\mathbf{r},
\mathbf{p}-\dfrac{\hbar \mathbf{k}}{2},t)+f(\mathbf{r},
\mathbf{p}+\dfrac{\hbar \mathbf{k}}{2},t)\Big]\Big\}=
$$\vspace{0.3cm}
$$
=\mathbf{A}(\mathbf{r},t)\dfrac{ie}{mc\hbar}\mathbf{p}\Big(f_+-f_-\Big),
$$
где
$$
f_{\pm}\equiv f(\mathbf{r},\mathbf{p}\mp\dfrac{\hbar
\mathbf{k}}{2},t).
$$

Следовательно, интеграл Вигнера---Власова равен:
$$
W[f]=\dfrac{ie}{mc\hbar}\mathbf{p}\mathbf{A}
(\mathbf{r},t)\Big[f_+^{}-f_-^{}\Big]=
\dfrac{iep_T}{mc\hbar}\mathbf{P}\mathbf{A}(\mathbf{r},t)
\Big[f_+^{}-f_-^{}\Big]=$$$$=\dfrac{iev_T}{c\hbar}\mathbf{P}\mathbf{A}
(\mathbf{r},t)\Big[f_+^{}-f_-^{}\Big].
\eqno{(3.3)}
$$

Здесь и ниже выражение $\mathbf{P} \mathbf{A}$ означает
скалярное произведение.

Далее удобнее использовать безразмерную скорость $\mathbf{C}$ в виде:
$$
\mathbf{C}=\dfrac{\mathbf{v}}{v_T}=\dfrac{\mathbf{p}}{p_T}-
\dfrac{e}{cp_T}\mathbf{A}(\mathbf{r},t)\equiv \mathbf{P}-
\dfrac{e}{cp_T}\mathbf{A}(\mathbf{r},t),
$$
где $\mathbf{P}=\dfrac{\mathbf{p}}{p_T}$ -- безразмерный
импульс.

В линейном приближении функцию $f$ в интеграле Вигнера---Вла\-со\-ва можно заменить на абсолютный максвеллиан, т.е. положить $f=f_M(P)$, где
$$
f_M(P)=N\Big(\dfrac{\beta}{\pi}\Big)^{3/2}e^{-P^2}.
$$
При этом интеграл Вигнера---Власова (3.3) будет иметь вид:
$$
W[f_M]=\dfrac{iev_T}{c\hbar}\mathbf{P}\mathbf{A}
(\mathbf{r},t)\Big[f_F^+-f_F^-\Big],
$$
где
$$
f_M^{\pm}\equiv f_M^{\pm}(\mathbf{P})=
N\Big(\dfrac{\beta}{\pi}\Big)^{3/2}e^{-\mathbf{P_{\pm}^2}}.
$$
а $p_T=mv_T$ -- тепловой импульс электронов.
Здесь
$$
P^2_{\pm}=\Big(\mathbf{P}\mp\dfrac{\hbar \mathbf{k}}{2p_T}\Big)^2
=\Big(\mathbf{P}\mp\dfrac{\mathbf{q}}{2}\Big)^2,\qquad \mathbf{q}=
\dfrac{\mathbf{k}}{k_T}.
$$
Линеаризацию равновесного распределения Максвелла---Больцмана
(3.2) проведем относительно
векторного потенциала $\mathbf{A}(\mathbf{r},t)$:
$$
f^{(0)}=f^{(0)}\Big|_{\mathbf{A}=0}+
\dfrac{\partial f^{(0)}}
{\partial \mathbf{A}}\Bigg|_{\mathbf{A}=0}\mathbf{A}(\mathbf{r},t),
$$
или, в явном виде:
$$
f^{(0)}=f_M(P)\Big(1+\dfrac{2e}{cp_T}\mathbf{P}\mathbf{A}(\mathbf{r},t)\Big).
\eqno{(3.4)}
$$
Учитывая разложение (3.4), ищем функцию Вигнера в виде:
$$
f=f_M(P)\Big[1+\dfrac{2e}{cp_T}\mathbf{P}\mathbf{A}(\mathbf{r},t)+
\mathbf{P}\mathbf{A}(\mathbf{r},t)h(\mathbf{P})\Big].
\eqno{(3.5)}
$$

Получаем следующее уравнение:
$$
[\mathbf{P}\mathbf{A}(\mathbf{r},t)]\;f_M(P)h(\mathbf{P})(1-i\omega\tau+
i\mathbf{k}_1\mathbf{P})=
$$\smallskip
$$
=\dfrac{iel_T}{c\hbar}
[\mathbf{P}\mathbf{A}(\mathbf{r},t)](f_F^+-f_F^-)+
[\mathbf{P}\mathbf{A}(\mathbf{r},t)]
\dfrac{2ie}{cp_T}f_M(P)(\omega\tau-l_T\mathbf{k}_1\mathbf{P}).
$$

Из этого уравнения находим:
$$
\big[\mathbf{P}\mathbf{A}(\mathbf{r},t)\big]f_M(P)h(\mathbf{P})=
\dfrac{2ie}{cp_T}\Xi(\mathbf{P},\mathbf{k}_1)
\big[\mathbf{P}\mathbf{A}(\mathbf{r},t)\big]
$$
где
$$
\Xi(\mathbf{P},\mathbf{k}_1)=
\dfrac{\omega \tau-l_T \mathbf{k}_1\mathbf{P}}
{1-i\omega\tau+il_T\mathbf{k}_1\mathbf{P}}f_M(P)+
\dfrac{\E_T}{\hbar \nu}
\dfrac{f_M^+-f_M^-}{1-i\omega\tau+il_T\mathbf{k}_1\mathbf{P}}.
\eqno{(3.6)}
$$

С помощью (3.5) и (3.6) построим полную функцию распределения:
$$
f=f^{(0)}+\dfrac{2ie}{cp_T}f_M(P)(\mathbf{PA})h(\mathbf{P})=
$$
$$
f=f^{(0)}+\dfrac{2ie}{cp_T}(\mathbf{PA})\Xi(\mathbf{P,k}).
\eqno{(3.7)}
$$ \medskip

Здесь $\mathbf{k}_1=\mathbf{k}l$, $l$ -- средняя длина свободного
пробега электронов, $\;l=v_T\tau$, $\mathbf{k}_1$ --
безразмерный волновой вектор.

\begin{center}
{\bf 4. ПЛОТНОСТЬ ЭЛЕКТРИЧЕСКОГО ТОКА}
\end{center}

Рассмотрим связь поля и потенциалов:
$$
\mathbf{E}(\mathbf{r},t)=
-\dfrac{1}{c}\dfrac{\partial \mathbf{A}(\mathbf{r},t)}{\partial t}-
\dfrac{\partial U(\mathbf{r},t)}{\partial \mathbf{r}},
$$
или
$$
\mathbf{E}(\mathbf{r},t)=\dfrac{i \omega}{c}\mathbf{A}(\mathbf{r},t).
$$

Следовательно, плотность тока связано с векторным потенциалом:
$$
\mathbf{j}(\mathbf{r},t)=
\sigma_{tr}\dfrac{i \omega}{c}\mathbf{A}(\mathbf{r},t).
$$

По определению, плотность тока равна:
$$
\mathbf{j}(\mathbf{r},t)=e\int \mathbf{v}(\mathbf{r},\mathbf{p},t)f
d^3v.
$$

Заметим, что плотность тока в равновесном состоянии (калибровочая
плотность тока) равна нулю:
$$
\mathbf{j}^{(0)}(\mathbf{r},t)=
e\int \mathbf{v}(\mathbf{r},\mathbf{p},t)f^{(0)}
d^3v=0.
$$

В самом деле, учитывая, что средняя скорость электронов в равновесном
состоянии равна нулю,  имеем:
$$
\mathbf{j}^{(0)}(\mathbf{r},t)=
e N^{(0)}\mathbf{u}^{(0)}(\mathbf{r},t)\equiv 0.
$$

Следовательно, с использованием равенства (3.8) для плотности тока
имеем равенство:
$$
\mathbf{j}(\mathbf{r},t)=
\dfrac{2ie^2}{cmp_T}\int \big(\mathbf{A\mathbf{P}}\big)
\mathbf{v}(\mathbf{r},\mathbf{p},t)\Xi(\mathbf{P,k})d^3v.
$$

Подставляя в это равенство явное выражение для скорости
$$
\mathbf{v}(\mathbf{r},\mathbf{P},t)=
\dfrac{\mathbf{p}}{m}-\dfrac{e \mathbf{A}(\mathbf{r},t)}{mc}=
\dfrac{p_T\mathbf{P}}{m}-\dfrac{e \mathbf{A}(\mathbf{r},t)}{mc},
$$
и, линеаризуя его по векторному полю, получаем:
$$
\mathbf{j}(\mathbf{r},t)=
\dfrac{2ie^2}{cm}\int
\big[\mathbf{P}\mathbf{A}(\mathbf{r},t)\big]\mathbf{P}\Xi(\mathbf{P,k})d^3v.
\eqno{(4.1)}
$$

Возьмем единичный вектор $\mathbf{e}_1=\dfrac{\mathbf{A}}{A}$,
направленный вдоль вектора $\mathbf{A}$. Тогда равенство (4.2)
можно записать в виде:
$$
\mathbf{j}(\mathbf{r},t)=\dfrac{2ie^2 A(\mathbf{r},t)}
{cm}
\int \big(\mathbf{P}\mathbf{e}_1\big)\mathbf{P}\Xi(\mathbf{P,k})\,d^3P,
\eqno{(4.2)}
$$

В силу симметрии значение интеграла (4.2) не изменится, если
вектор $\mathbf{e}_1$ заменить на любой другой единичный вектор
$\mathbf{e}_2$, перпендикулярный вектору $\mathbf{k}_1$, т.е.
$$\qquad \qquad \mathbf{e}_2=\dfrac{\mathbf{A} \times \mathbf{k}_1}
{|\mathbf{A} \times \mathbf{k}_1|}=
\dfrac{\mathbf{A} \times \mathbf{k}_1}{Ak_1},
$$
причем $\mathbf{A} \times \mathbf{k}_1$ есть векторное произведение.

Разложим вектор $\mathbf{P}$ по трем ортогональным направлениям
$\mathbf{e}_1$, $\mathbf{e}_2$ и
$\mathbf{n}=\dfrac{\mathbf{k}_1}{k_1}$:
$$
\mathbf{P}=(\mathbf{Pn})\mathbf{n}+(\mathbf{P}\mathbf{e}_1)
\mathbf{e}_1+(\mathbf{P}\mathbf{e}_2)\mathbf{e}_2.
$$

С помощью этого разложения получаем:
$$
(\mathbf{PA})\mathbf{P}=A(\mathbf{P}\mathbf{e}_1)\mathbf{P}=
$$
$$
=A(\mathbf{P}\mathbf{e}_1)(\mathbf{Pn})\mathbf{n}+
A(\mathbf{P}\mathbf{e}_1)^2\mathbf{e}_1+A(\mathbf{P}\mathbf{e}_1)
(\mathbf{P}\mathbf{e}_2)\mathbf{e}_2.
$$

Подставляя это разложение в (4.2), и, учитывая, что интегралы
от нечетных функций по симметричному промежутку равны нулю, получаем:
$$
\mathbf{j}(\mathbf{r},t)=\dfrac{2ie^2\mathbf{A}(\mathbf{r},t)}
{cm}\int (\mathbf{P}\mathbf{e}_1)^2
\Xi(P^2,\mathbf{Pn})\,d^3P.
\eqno{(4.3)}
$$

Ввиду симметрии значение интеграла не изменится,
если вектор $\mathbf{e}_1$ заменить на любой другой единичный
вектор $\mathbf{e}_2$, перпендикулярный вектору $\mathbf{k}_1$. Поэтому
$$
\int \big(\mathbf{e}_1\mathbf{P}\big)^2[\cdots]d^3P=
\int \big(\mathbf{e}_2\mathbf{P}\big)^2[\cdots]d^3P=
$$
$$
=\dfrac{1}{2}\int \Big[\Big(\mathbf{e}_1 \mathbf{P}\Big)^2
+\Big(\mathbf{e}_2\mathbf{P}\Big)^2\Big]
[\cdots]d^3P.
$$

Заметим, что квадрат длины вектора $\mathbf{P}$ равен:
$$
P^2=(\mathbf{P}\mathbf{e}_1)^2+(\mathbf{P}\mathbf{e}_2)^2+
(\mathbf{Pn})^2,
$$
откуда
$$
\big(\mathbf{e}_1\mathbf{P}\big)^2
+\big(\mathbf{e}_2\mathbf{P}\big)^2=P^2-
\dfrac{(\mathbf{P}\mathbf{k}_1)^2}{k_1^2}=P^2-(\mathbf{Pn})^2=
P_\perp^2,
$$
где $P_\perp$ -- проекция вектора $\mathbf{P}$ на прямую,
перпендикулярную плоскости $(\mathbf{e}_1, \mathbf{e}_2)$.

Отсюда для плотности тока получаем следующее выражение
$$
\mathbf{j}(\mathbf{r},t)=
\dfrac{ie^2\mathbf{A}(\mathbf{r},t)}{cm}\int \Xi(\mathbf{P,k})P_\perp^2 d^3v.
$$

Заменяя ток в левой части этого равенства выражением через поле,
получаем:
$$
\sigma_{tr}=
\dfrac{e^2}{m\omega}\int \Xi(\mathbf{P,k})P^2_\perp d^3v.
$$\medskip

\begin{center}
  \bf 5. ЭЛЕКТРИЧЕСКАЯ ПРОВОДИМОСТЬ И ДИЭЛЕКТРИЧЕСКАЯ ПРОНИЦАЕМОСТЬ
\end{center}

Запишем предыдущее выражение в явном виде:
$$
\sigma_{tr}=\dfrac{e^2N}{m\omega \pi^{3/2}}\int \Bigg[\dfrac{\omega\tau-\mathbf{k}_1\mathbf{P}}{1-i\omega\tau+i
\mathbf{k}_1\mathbf{P}}f_0(P)+\dfrac{\E_T}{\hbar \nu}\dfrac{f_0^+-f_0^-}{1-i\omega\tau+i\mathbf{k}_1\mathbf{P}}\Bigg]
P_\perp^2d^3P.
\eqno{(5.1)}
$$

Здесь
$$
f_0(P)=e^{-P^2},\qquad f_0^{\pm}=f_0(\mathbf{P}_{\pm})=e^{-\mathbf{P}_{\pm}^2}=
e^{-\Big(\mathbf{P}\mp\dfrac{\mathbf{q}}{2}\Big)^2}.
$$

Из последней формулы получаем выражение поперечной
диэлектрической проницаемости в квантовой максвелловской плазме:
$$
\sigma_{tr}=\sigma_0\dfrac{\nu}{\omega \pi^{3/2}}\int
\Xi(\mathbf{P,k})P_\perp^2d^3P.
\eqno{(5.2)}
$$
Здесь $\sigma_0=\dfrac{e^2N}{m \nu}$ -- классическая проводимость.

Диэлектрическая проницаемость квантовой максвелловской столкновительной плазмы определяется выражением:
$$
\varepsilon_{tr}=1+\dfrac{4\pi i}{\omega}\sigma_{tr}=
1+i\dfrac{\omega_p^2}{\omega^2}\dfrac{1}{\pi^{3/2}}\int
\Xi(\mathbf{P,k})P_\perp^2 d^3P.
\eqno{(5.3)}
$$

В длинноволновом пределе (когда $k\to 0$) из формулы (5.2) получаем
известное классическое выражение:
$$
\sigma_{tr}=\dfrac{\sigma_0}{1-i\omega\tau}=\sigma_0\dfrac{\nu}{\nu-i\omega}.
$$

Рассмотрим квантовомеханический предел ($\hbar\to 0$) электрической
проводимости в случае произвольных значений волнового числа, т.е.
предел электрической проводимости в случае, когда постоянная Планка
стремится к нулю, а величина волнового числа $k$ -- произвольная.
При малых $\hbar$ имеем:
$$
f_0^{\pm}=e^{-P^2\pm \mathbf{P}\dfrac{\mathbf{k}}{k_T}}=
f_0(P)\Big(1\pm \mathbf{P}\dfrac{\mathbf{k}}{k_T}\Big)=
$$
$$
=f_0(P)\Big(1\pm\dfrac{P_xk}{k_T}\Big)=f_0(P)
\Big(1\pm \dfrac{\hbar P_x k}{mv_T}\Big).
$$

Следовательно,
$$
(\omega\tau-\mathbf{k}l_T\mathbf{P})f_0(P)+\dfrac{p_T^2\tau}{2m\hbar}
[f_0(\mathbf{P}_+-f_0(\mathbf{P}_-)]=\omega\tau f_0(P).
$$

Таким образом, в линейном приближении при малых $\hbar$
(независимо от величины $k$) для поперечной электрической
проводимости максвелловской квантовой столкновительной плазмы получаем:
$$
\sigma_{tr}=\sigma_{tr}^{\rm classic},
\eqno{(5.4)}
$$
где
$$
\dfrac{\sigma_{tr}^{\rm classic}}{\sigma_0}=\dfrac{1}{\pi^{3/2}}\int
\dfrac{e^{-P^2}P_\perp^2 d^3P}{1-i\omega\tau+i \mathbf{k}_1\mathbf{P}}.
\eqno{(5.5)}
$$
Выражение (5.5) в точности совпадает с выражением поперечной электрической
проводимости для классической максвелловской плазмы с произвольной температурой.

Вернемся к выражению (5.2). Представим его в виде суммы двух слагаемых:
$$
\sigma_{tr}=\sigma_{tr}^{\rm classic}+\sigma_{tr}^{\rm quant},
\eqno{(5.6)}
$$
где первое слагаемое определяется равенством (5.5), а второе -- равенством:
$$
\dfrac{\sigma_{tr}^{\rm quant}}{\sigma_0}=\dfrac{1}{\pi^{3/2}}\int
\Bigg[-\dfrac{\mathbf{k}_1\mathbf{P}}{\omega\tau}+
\dfrac{\E_T}{\hbar \omega}[f_0^+-f_0^-]\Bigg]
\dfrac{P_\perp^2d^3P}{1-i\omega\tau+i \mathbf{k}_1\mathbf{P}}
\eqno{(5.7)}
$$

Точно так же для диэлектрической проницаемости можно написать:
$$
\varepsilon_{tr}=\varepsilon_{tr}^{\rm classic}+\varepsilon_{tr}^{\rm quant}.
\eqno{(5.7)}
$$

В (5.7)
$$
\varepsilon_{tr}^{\rm classic}=1+i\dfrac{\omega_p^2}{\omega^2}\omega\tau\dfrac{1}{\pi^{3/2}}\int
\dfrac{P_\perp^2d^3P}{1-i\omega\tau+i \mathbf{k}_1\mathbf{P}},
\eqno{(5.8)}
$$
$$
\varepsilon_{tr}^{\rm quant}=i\dfrac{\omega_p^2}{\omega^2}\dfrac{1}{\pi^{3/2}}\int
\Bigg[-\mathbf{k}_1\mathbf{P}e^{-P^2}+\dfrac{\E_T}{\hbar \nu}
\Big(e^{-\mathbf{P_+}^2}-e^{-\mathbf{P_-^2}}\Big)\Bigg]
\dfrac{P_\perp^2d^3P}{1-i\omega\tau+i \mathbf{k}_1\mathbf{P}}
\eqno{(5.9)}
$$

Выражение (5.8) в точности совпадает с классическим выражением диэлектрической проницаемост максвелловской плазмы, а выражение (5.9) определяется квантовыми свойствами максвелловской плазмы.

Согласно (5.7) получаем выражение для
диэлектрической проницаемости в квантовой столкновительной плазме:
$$
\varepsilon_{tr}=1+\dfrac{\omega_p^2}{\omega^2}
 \dfrac{i}{4\pi f_2(\alpha)}
\int\Bigg\{\Big[\omega\tau-\mathbf{k}_1\mathbf{P}\Big]g(P)+
\hspace{4cm}
$$
$$\hspace{4cm}
+\dfrac{\E_T}{\hbar
\nu}\Big[f_F^+(\mathbf{P})-f_F^-(\mathbf{P})\Big]\Bigg\}
\dfrac{P_\perp^2\;d^3P}{1-i\omega \tau+i
\mathbf{k}_1\mathbf{P}}.
\eqno{(5.10)}
$$

\begin{center}
{\bf 6. ПРАВИЛО СУММ}
\end{center}

Проверим выполнение одного из соотношений, называемого
 называемого правилом $f$-сумм
(см., например, \cite{Pains},\cite{Dressel}) для поперечной диэлектрической
проницаемости
(6.14). Это правило выражается формулой (4.200) из монографии \cite{Pains}
$$
\int\limits_{-\infty}^{\infty}\varepsilon_{tr}(q,\omega,\nu)\omega d\omega=
\pi \omega_p^2.
\eqno{(6.1)}
$$

Как показано в \cite{Pains}, для доказательства соотношения (6.1) достаточно
доказать выполнение предельного соотношения
$$
\varepsilon_{tr}(q,\omega,\nu)=1-\dfrac{\omega_p^2}{\omega^2}+
o\Big(\dfrac{1}{\omega^2}\Big),\qquad \omega\to \infty.
\eqno{(6.2)}
$$
Из выражения (5.9) видно, что
$$
\varepsilon_{tr}^{\rm quant}=o\Big(\dfrac{1}{\omega^2}\Big),\qquad
\omega\to\infty.
$$
С помощью этого предельного соотношения представим соотношение (5.10) с
использованием равенства (5.7) при больших $\omega$ ($\omega \gg 1$) в виде:
$$
\varepsilon_{tr}=1+\dfrac{i\omega_p^2 \omega \tau}{\pi^{3/2}\omega^2}
\int \dfrac{e^{-P^2}P_\perp^2\;d^3P}{1-i\omega \tau+i\mathbf{k}_1\mathbf{P}}+
o\Big(\dfrac{1}{\omega^2}\Big),\qquad \omega\to \infty.
\eqno{(6.3)}
$$

Нетрудно видеть, что соотношение (6.3) при $\omega\to \infty$ записывается в
виде
$$
\varepsilon_{tr}=1-\dfrac{\omega_p^2}{\omega^2\pi^{3/2}}\int
e^{-P^2}P_\perp^2 d^3P+o\Big(\dfrac{1}{\omega^2}\Big), \qquad \omega\to \infty.
\eqno{(6.4)}
$$

Из соотношения (6.4) и вытекает соотношение (6.2), если учесть, что
$$
\dfrac{1}{\pi^{3/2}}\int e^{-P^2}P_\perp^2 d^3P=1.
$$

\begin{center}
{\bf 7. ВЫЧИСЛЕНИЕ ЭЛЕКТРИЧЕСКОЙ ПРОВОДИМОСТИ И ДИЭЛЕКТРИЧЕСКОЙ
ПРОНИЦАЕМОСТИ}
\end{center}

Начнем с электрической проводимости. Вектор $\mathbf{k}$ направим вдоль
оси $x$. Заметим, что
$$
1-i\omega\tau+iklP_x=ikl\Big(P_x-\dfrac{\omega+i \nu}{k\nu l}\Big)=
ikl\Big(P_x-\dfrac{\omega+i \nu}{k\nu l}\Big)=$$$$=
ikl\Big(P_x-\dfrac{z}{q}\Big)=i\dfrac{q}{y}\Big(P_x-\dfrac{z}{q}\Big),
$$ где $q$ -- безразмерное волновое число, $q=\dfrac{k}{k_T}$,
$$
z=x+iy=\dfrac{\omega+i \nu}{k_Tv_T} , \qquad x=\dfrac{\omega}{k_Tv_T},\qquad
y=\dfrac{\nu}{k_Tv_T}.
$$
Заметим также, что
$$
\mathbf{P}_{\pm}^2=\Big(\mathbf{P}\mp \dfrac{\hbar \mathbf{k}}{2p_T}\Big)^2=
\Big(\mathbf{P}-\dfrac{\mathbf{q}}{2}\Big)^2=$$$$=
\Big(P_x-\dfrac{q}{2}\Big)^2+P_y^2+P_z^2=\Big(P_x-\dfrac{q}{2}\Big)^2+P_\perp^2.
$$
Теперь формулу (5.1) можно переписать в виде:
$$
\dfrac{\sigma_{tr}}{\sigma_0}=
\dfrac{1}{iqk_Tl_T\pi^{3/2}}\int
\dfrac{e^{-P^2}P_\perp^2d^3P}
{P_x-z/q}.
$$
Эту формулу  представим в виде 
$$
\dfrac{\sigma_{tr}^{\rm classic}}{\sigma_0}=-\dfrac{iy}{q}\dfrac{1}{\sqrt{\pi}}
\int\limits_{-\infty}^{\infty}\dfrac{e^{-\tau^2}d\tau}{\tau-z/q}=-\dfrac{iy}{q}t(z/q),
\eqno{(7.1)}
$$
где
$$
t(z/q)= \dfrac{1}{\sqrt{\pi}}\int\limits_{-\infty}^{\infty}
\dfrac{e^{-\tau^2}d\tau}{\tau-z/q}.
$$

Точно так же, для классической диэлектрической функции получаем:
$$
\varepsilon_{tr}^{\rm classic}=1+\dfrac{x_p^2}{xq\sqrt{\pi}}
\int\limits_{-\infty}^{\infty}\dfrac{e^{-\tau^2}d\tau}{\tau-z/q}=1+\dfrac{x_p^2}{xq}t(z/q).
\eqno{(7.2)}
$$
Здесь
$$
\dfrac{\omega_p^2}{\omega^2}=\dfrac{x_p^2}{x^2},\qquad x_p=\dfrac{\omega_p}
{k_Tv_T}.
$$

Перейдем к вычислению квантовых поправок. Для электрической проводимости
имеем:
$$
\dfrac{\sigma_{tr}^{\rm quant}}{\sigma_0}=
\dfrac{i \nu}{\omega\pi^{3/2}}\int\dfrac{e^{-P^2}P_xP_\perp^2d^3P}{P_x-z/q}-
$$$$\hspace{5cm}-\dfrac{i\E_T}{\hbar \omega k_Tl_Tq\pi^{3/2}}\int
\dfrac{(f_0^+-f_0^-)P_\perp^2d^3P}{P_x-z/q}.
$$
Отсюда получаем
$$
\dfrac{\sigma_{tr}^{\rm quant}}{\sigma_0}=\dfrac{iy}{x}\Big[\lambda(z/q)+
\dfrac{1}{2}T(z,q)\Big],
\eqno{(7.3)}
$$
где
$$
\lambda(z)=1+zt(z),\qquad
T(z,q)=\dfrac{1}{\sqrt{\pi}}\int\limits_{-\infty}^{\infty}
\dfrac{e^{-\tau^2}d\tau}{(\tau-z/q)^2-(q/2)^2}.
$$

На основании (7.3) для диэлектрической функции получаем:
$$
\varepsilon_{tr}^{\rm quant}=
-\dfrac{\omega_p}{\omega^2}[\lambda(z/q)+\dfrac{1}{2}
T(z,q)].
\eqno{(7.4)}
$$

Выпишем на основании предыдущих равенств формулу для электрической проводимости
$$
\dfrac{\sigma_{tr}}{\sigma_0}=\dfrac{iy}{x}\Big[-\dfrac{x}{q}t(z/q)+
\lambda(z/q)+\dfrac{1}{2}T(z,q)\Big]
\eqno{(7.5)}
$$
и для диэлектрической функции
$$
\varepsilon_{tr}=1-\dfrac{x_p^2}{x^2}\Big[-\dfrac{x}{q}t(z/q)+
\lambda(z/q)+\dfrac{1}{2}T(z,q)\Big].
\eqno{(7.6)}
$$
\begin{center}
\bf 8. АНАЛИЗ РЕЗУЛЬТАТОВ И ЗАКЛЮЧЕНИЕ
\end{center}

Для численного и графического анализа воспользуемся формулами (7.1)--(7.6).

На рис. 1--4 и 7,8 все кривые $1$ отвечают классической плазме, а кривые
$2$ -- квантовой.

На рис. 1 и 2 представлены зависимости действительной и мнимой частей
относительной электрической проводимости $\Re(\sigma_{tr}/\sigma_0)$
и $Im(\sigma_{tr}/\sigma_0)$  квантовой и классической плазмы 
от величины безразмерной частоты
колебаний векторного потенциала $x=\omega/k_Tv_T$ для случая $q=0.05,
y=0.01$. Отметим, что при $x \geqslant 0.5$ величины действительной и
мнимой частей классической и квантовой плазмы практически совпадают. 
При умнеьшении значения безразмерного волнового числа $q$ величины и
действительной и мнимой частей для классической и квантовой плазмы начинают
мало различаться и при $q=0.1$ практически совпадают. 
На рис. 3 и 4 представлены зависимости действительной и мнимой частей
относительной электрической проводимости классической и квантовой плазмы как 
функции безразмерного волнового числа для случая $x=1, y=0.01$. При 
$q \leqslant 1$ значения проводимостей  классической и квантовой плазмы
практически совпадают и при $q\to 0$ проводимость квантовой плазмы переходит
в проводимость классической. Отметим, что мнимая часть относительной 
проводимости квантовой плазмы стремится не к нулю при $q\to \infty$, как 
действительная часть, а стремится к конечному пределу. Найдем этот предел
$$
\lim\limits_{q\to \infty}\dfrac{\sigma_{tr}^{\rm quant}}{\sigma_0}=
\dfrac{iy}{x}.
$$

На рс. 5 и 6 представлены зависимости   действительной и мнимой частей
относительной электрической проводимости квантовой плазмы как
функции безразмерного волнового числа при различных значениях
безразмерной частоты колебаний векторного потенциала при $x=0.1,0.2,0.3$
(действительная часть) и при $x=1,1.5,2$ (мнимая часть). Заметим, что 
действительная часть проводимости имеет только максимум, а мнимая часть ---
и максимум, и минимум. С ростом величины $x$ значения и действительной,
и мнимой частей уменьшаются.

В настоящей работе выведены формулы для вычисления
поперечной электрической проводимости и диэлектрической проницаемости в
квантовой невырожденной
столкновительной плазме. Для этого используется кинетическое
уравнение Вигнера---Власова---Больцмана с интегралом
столкновений в форме БГК--модели (Бхатнагар, Гросс и Крук).
\begin{figure}[h]
\begin{center}
\includegraphics[width=16cm, height=8cm]{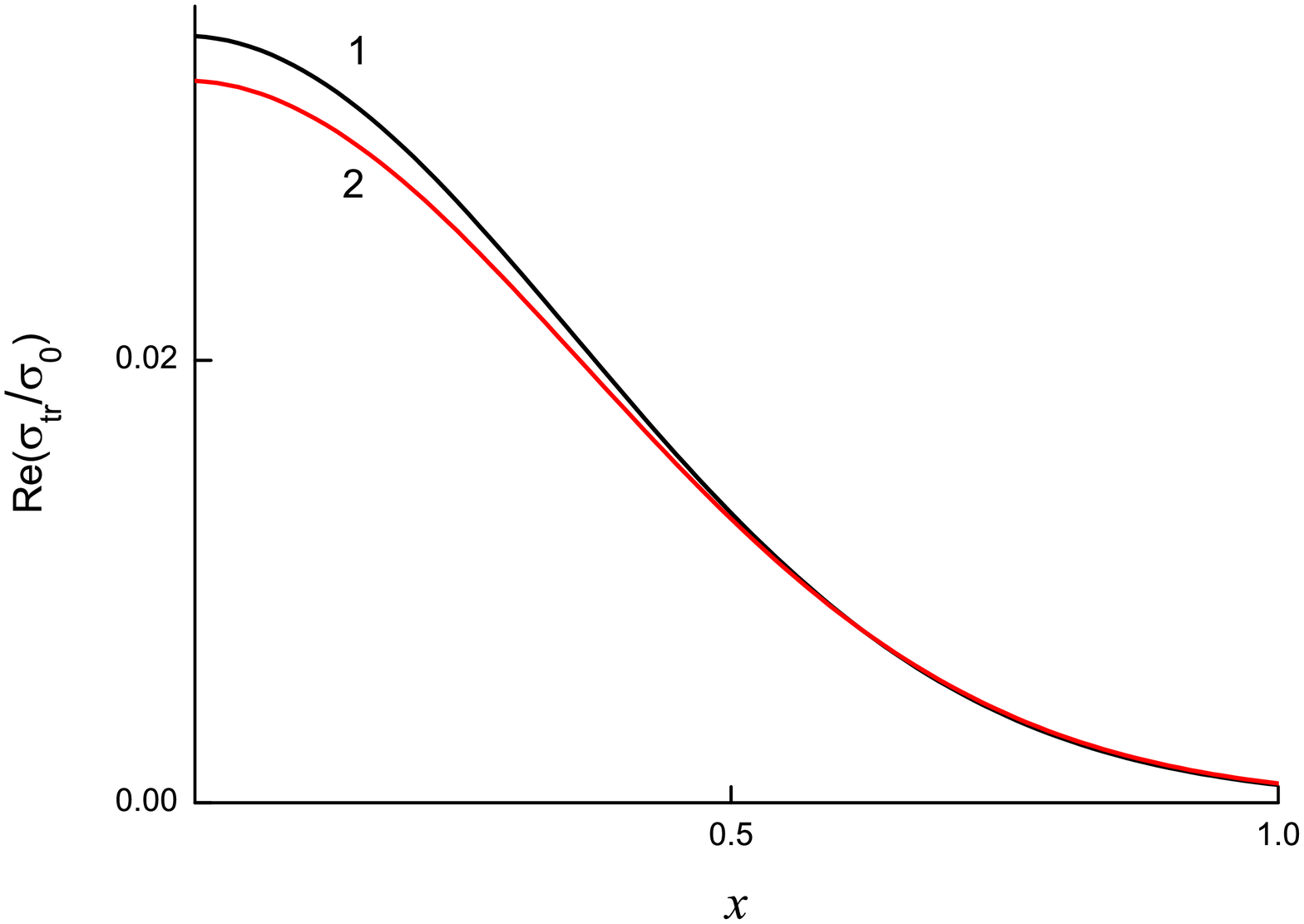}
\caption{Случай: $q=0.5, y=0.01.$
Зависимость $\Re(\sigma_{tr}/\sigma_0)$ от величины $x$.}
\end{center}
\begin{center}
\includegraphics[width=16cm, height=8cm]{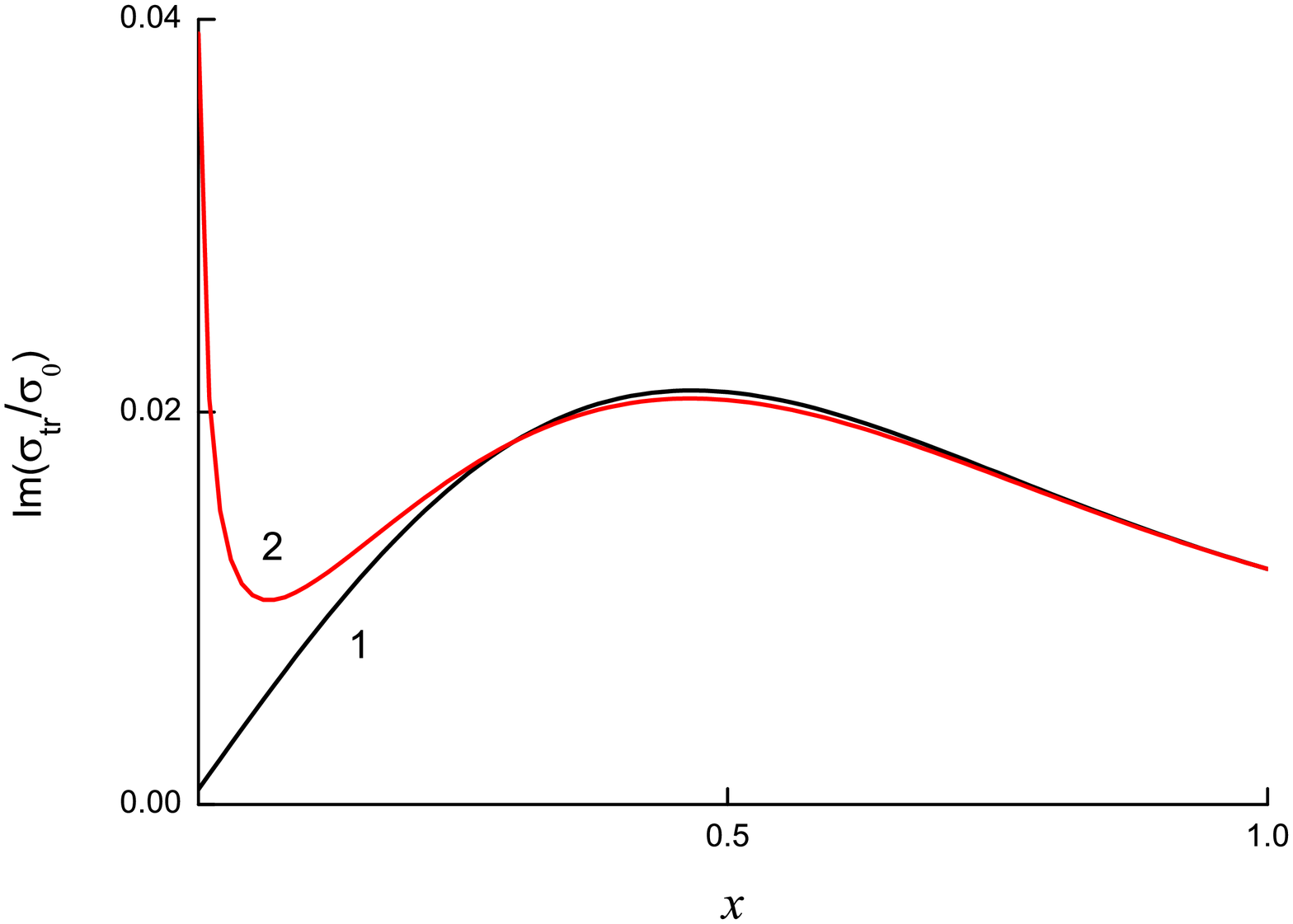}
\caption{Случай: $q=0.5, y=0.01.$
Зависимость $\Im(\sigma_{tr}/\sigma_0)$ от величины $x$.}
\end{center}
\end{figure}

\begin{figure}[t]
\begin{center}
\includegraphics[width=16cm, height=10cm]{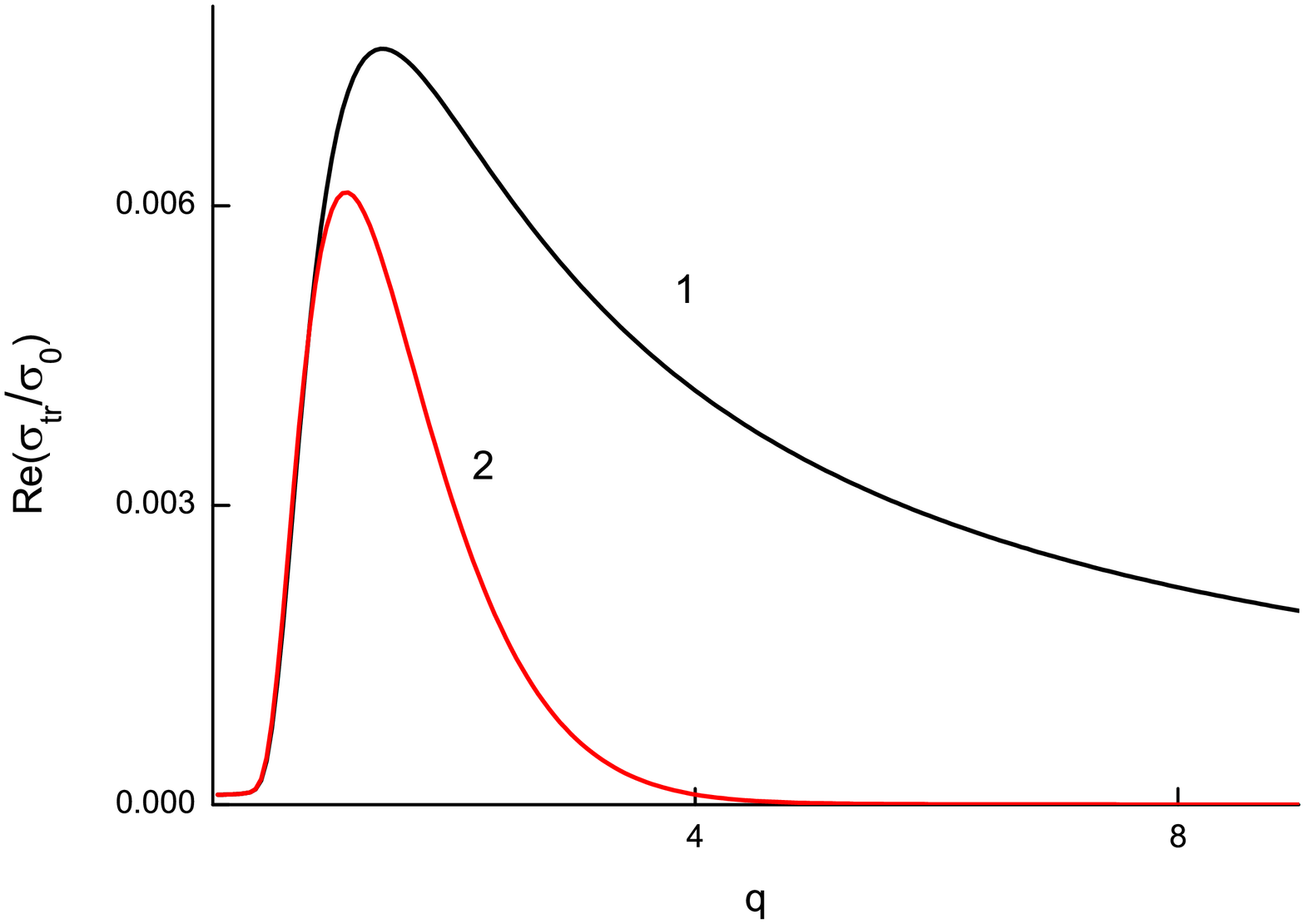}
\caption{Случай: $x=1, y=0.01.$
Зависимость $\Re(\sigma_{tr}/\sigma_0)$ от величины $q$.}
\end{center}
\begin{center}
\includegraphics[width=16cm, height=10cm]{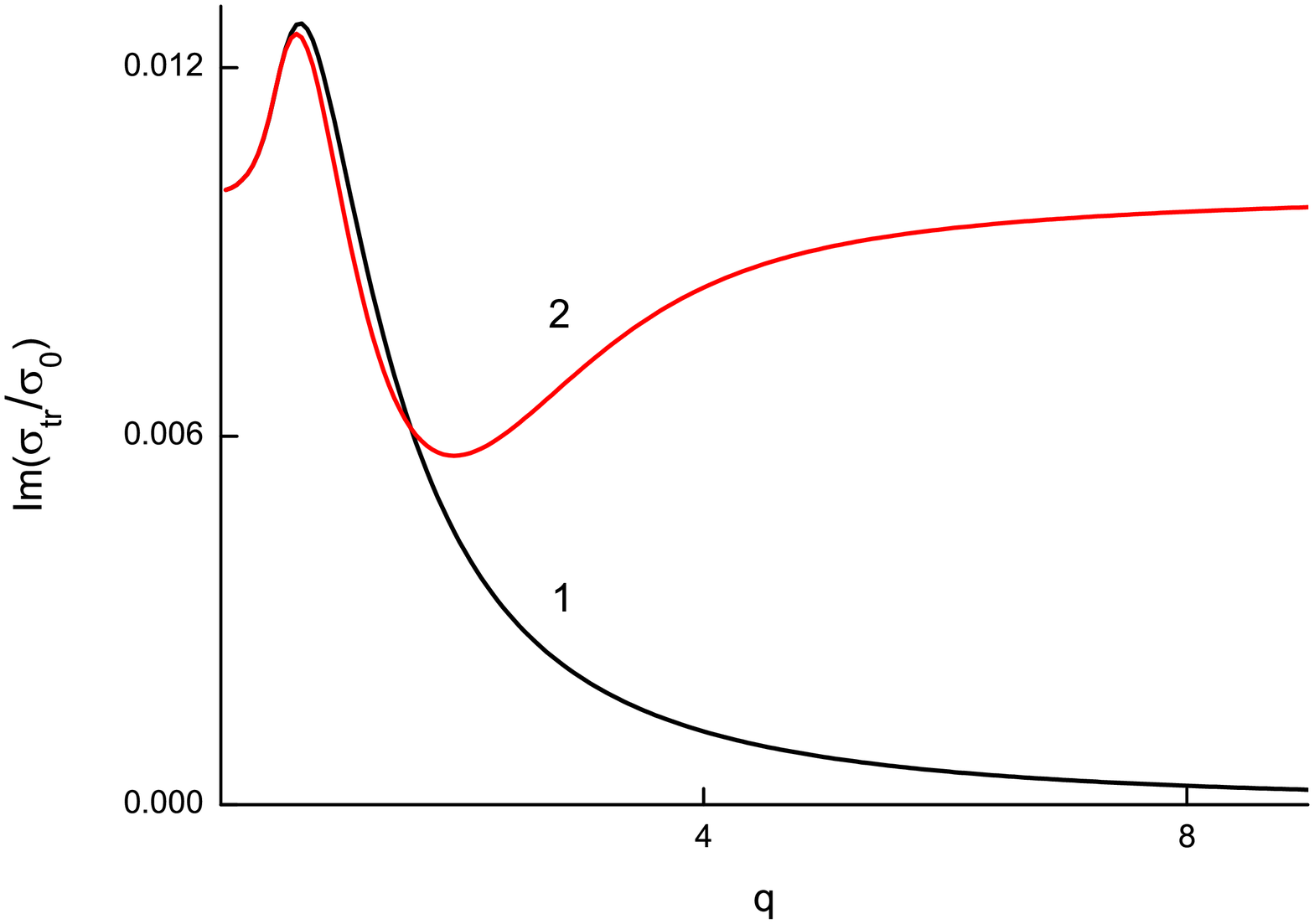}
\caption{Случай: $x=1, y=0.01.$
Зависимость $\Im(\sigma_{tr}/\sigma_0)$ от величины $q$.}
\end{center}
\end{figure}

\begin{figure}[t]
\begin{center}
\includegraphics[width=16cm, height=10cm]{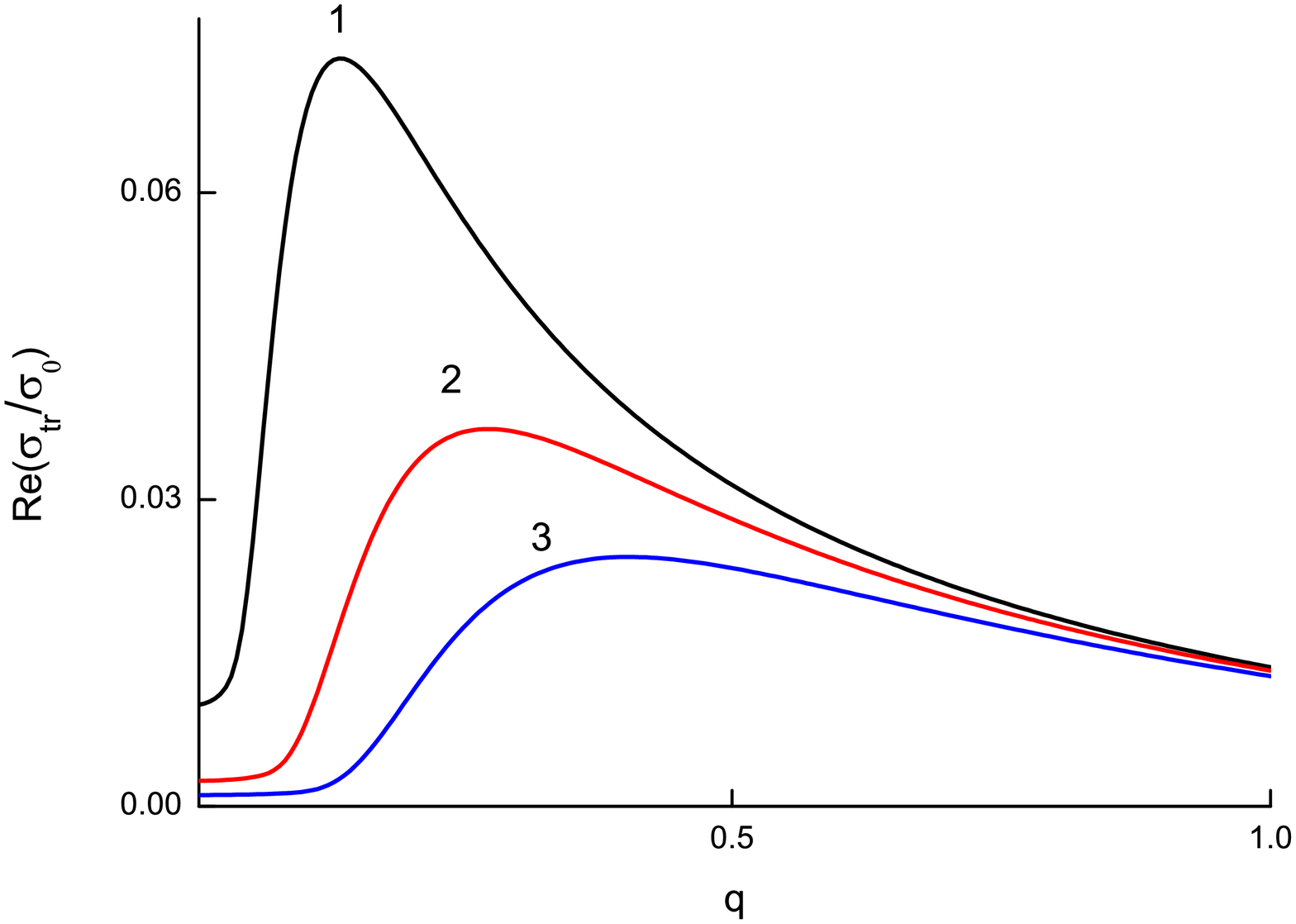}
\caption{Зависимость $\Re(\sigma_{tr}/\sigma_0)$ от величины $q$.
Кривые $1,2,3$ отвечают значениям параметра $x=0.1,0.2,0.3$. }
\end{center}
\begin{center}
\includegraphics[width=16cm, height=10cm]{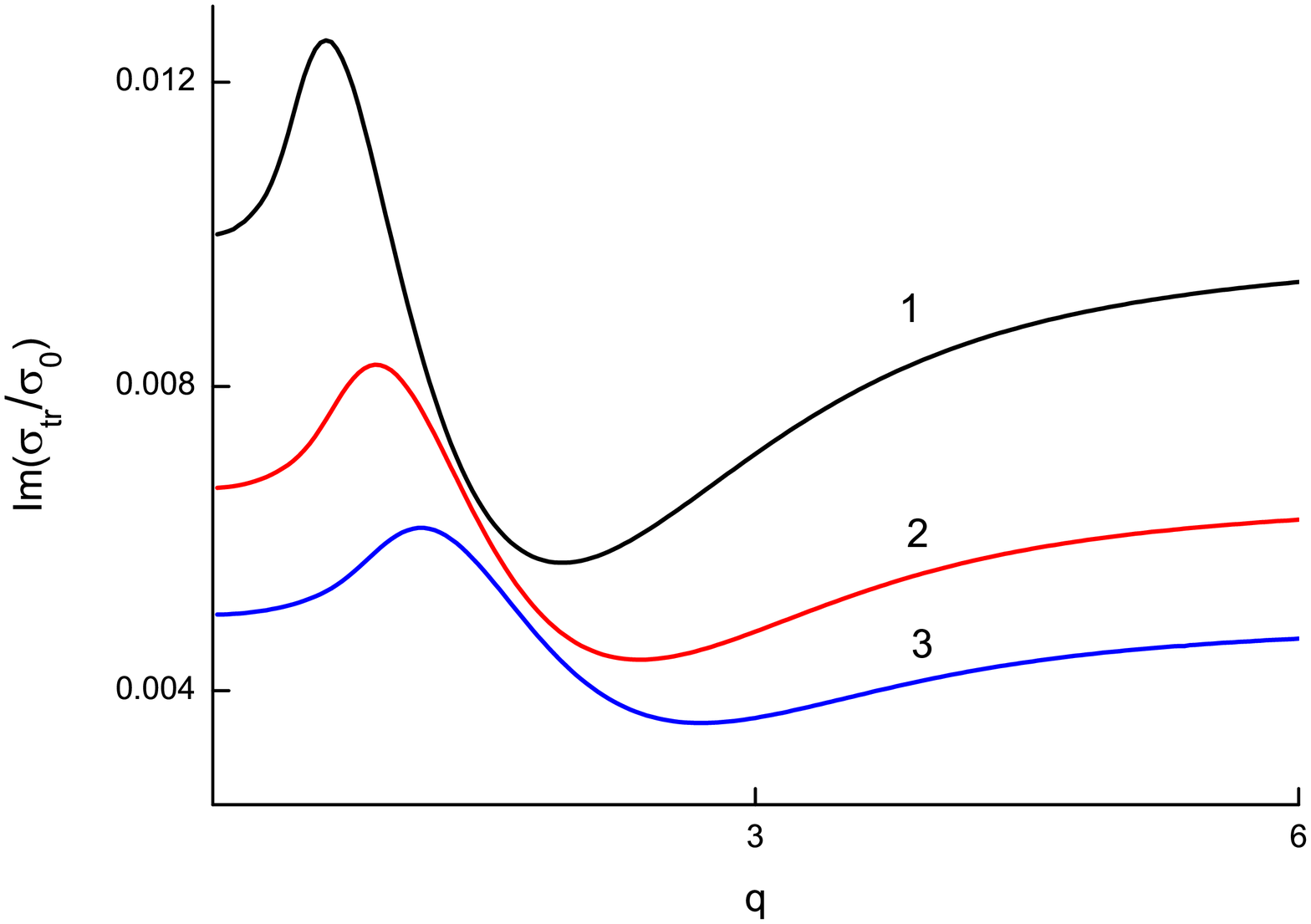}
\caption{Зависимость $\Im(\sigma_{tr}/\sigma_0)$ от величины $q$.
Кривые $1,2,3$ отвечают значениям параметра $x=1,1.5,2$.}
\end{center}
\end{figure}

\begin{figure}[h]
\begin{center}
\includegraphics[width=16cm, height=10cm]{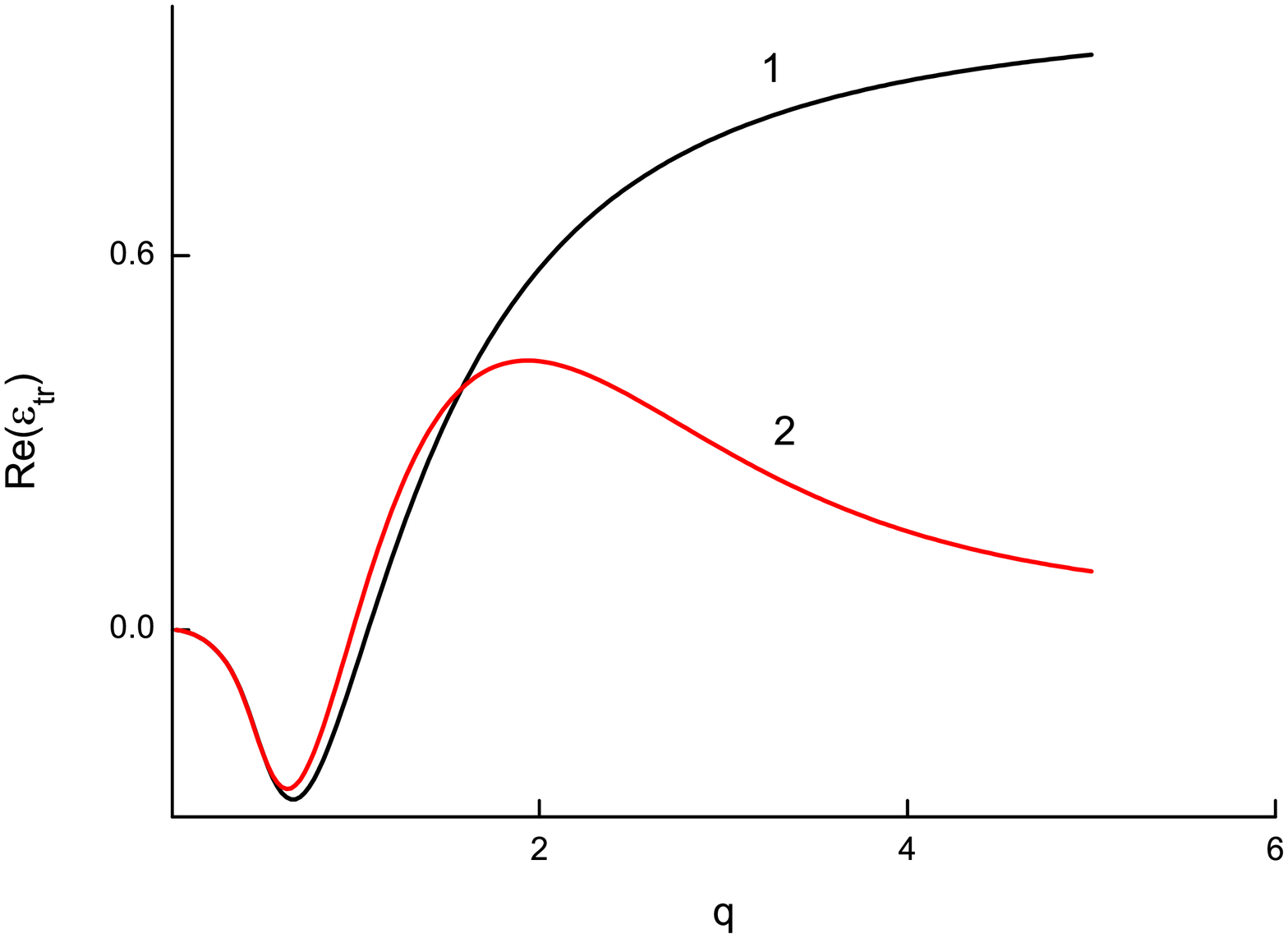}
\caption{Зависимость $\Re(\varepsilon_{tr})$ от величины $q$.}
\end{center}
\begin{center}
\includegraphics[width=16cm, height=10cm]{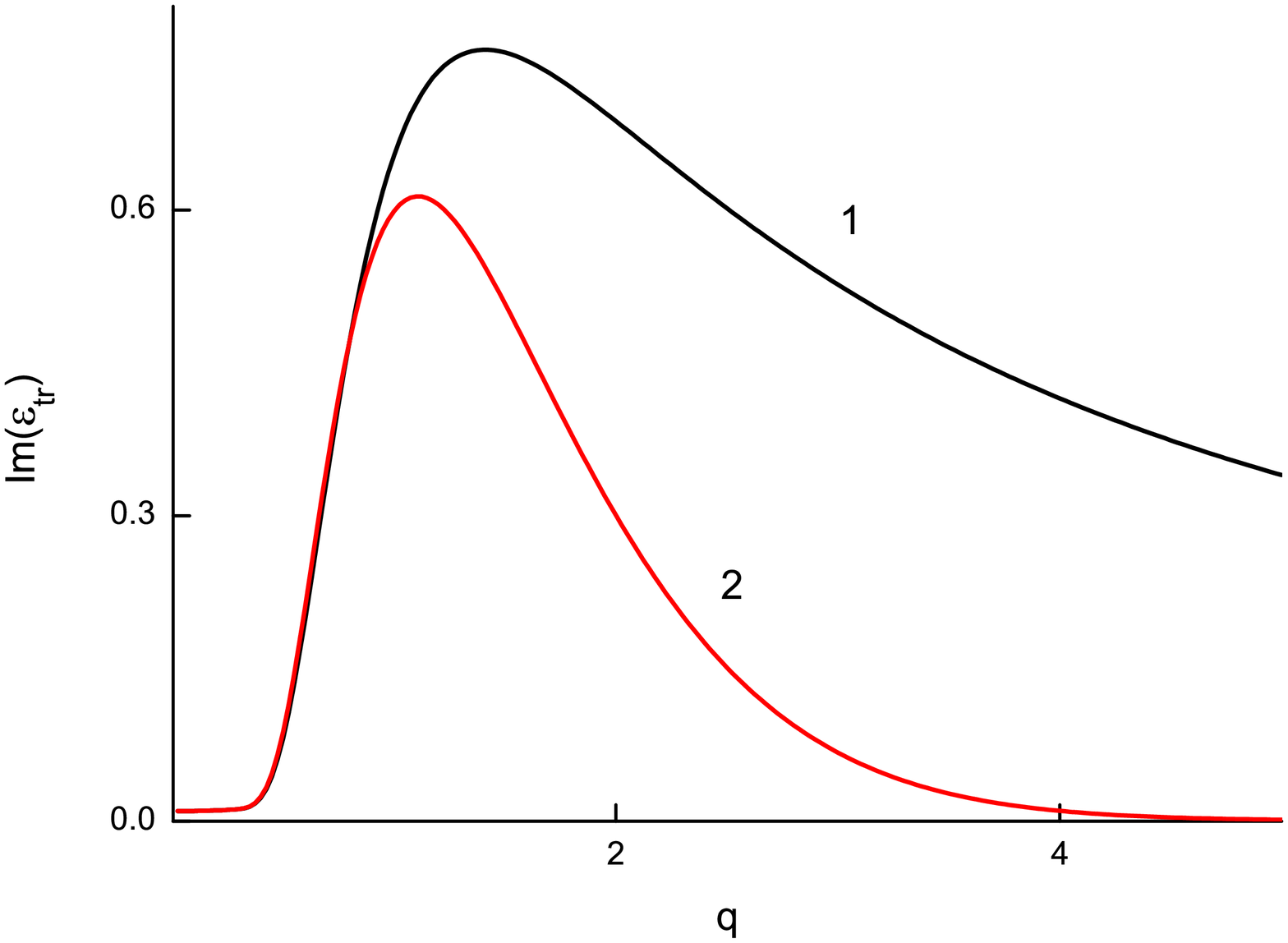}
\caption{Зависимость $\Im(\varepsilon_{tr})$ от величины $q$.}
\end{center}
\end{figure}

\end{document}